\documentclass{ws-rv975x65}
\usepackage{subfigure}     
\usepackage{ws-rv-thm}     
\usepackage[square]{ws-rv-van}
\makeindex
\renewcommand{\v}[1]{\boldsymbol{#1}}		

\begin{document}

\chapter[Manifestations of dark matter and variations of fundamental constants in atoms and astrophysical phenomena]{Manifestations of dark matter and variations of fundamental constants in atoms and astrophysical phenomena}\label{ra_ch1}

\author[Y.~V.~Stadnik and V.~V.~Flambaum]{Yevgeny V.~Stadnik$^{1}$ and Victor V.~Flambaum$^{1,2}$}

\address{$^{1}$School of Physics, University of New South Wales, Sydney 2052, Australia \\
y.stadnik@unsw.edu.au\\
v.flambaum@unsw.edu.au}
\address{$^{2}$Mainz Institute for Theoretical Physics, Johannes Gutenberg University Mainz, D 55122 Mainz, Germany}

\begin{abstract}
We present an overview of recent developments in the detection of light bosonic dark matter, including axion, pseudoscalar axion-like and scalar dark matter, which form either a coherently oscillating classical field or topological defects (solitons). We emphasise new high-precision laboratory and astrophysical measurements, in which the sought effects are linear in the underlying interaction strength between dark matter and ordinary matter, in contrast to traditional detection schemes for dark matter, where the effects are quadratic or higher order in the underlying interaction parameters and are extremely small. New terrestrial experiments include measurements with atomic clocks, spectroscopy, atomic and solid-state magnetometry, torsion pendula, ultracold neutrons, and laser interferometry. New astrophysical observations include pulsar timing, cosmic radiation lensing, Big Bang nucleosynthesis and cosmic microwave background measurements. We also discuss various recently proposed mechanisms for the induction of slow `drifts', oscillating variations and transient-in-time variations of the fundamental constants of Nature by dark matter, which offer a more natural means of producing a cosmological evolution of the fundamental constants compared with traditional dark energy-type theories, which invoke a (nearly) massless underlying field. 
Thus, measurements of variation of the fundamental constants gives us a new tool in dark matter searches.

\end{abstract}
\body

\section{Introduction}
\label{Intro}
\begin{figure}[h!]
\begin{center}
\includegraphics[width=10cm]{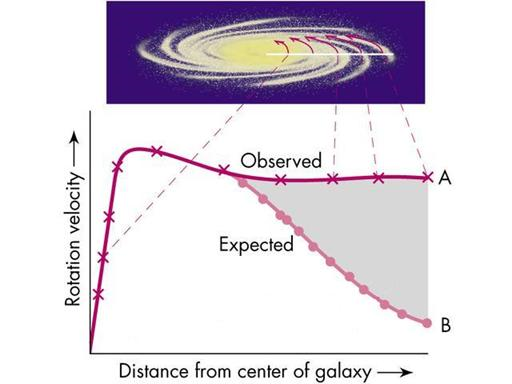}
\caption{Observed (purple line) and expected (pink line, which has the Kepplerian dependence $v \propto 1/\sqrt{r}$ at large distances) orbital velocities of stars as a function of distance from the galactic centre. The discrepancy between the two radial functions is consistent with the presence of dark matter haloes in galaxies.} 
\label{fig:DM_rotation_curve}
\end{center}
\end{figure}

Dark matter (DM) remains one of the most important unsolved problems in contemporary physics. Observations of stellar orbits about galactic centres from as early as the 1930s \cite{Zwicky1933,Zwicky1937}, which were later refined in the 1970s \cite{Rubin1970,Rubin1980}, have indicated that the orbital velocities of stars remain approximately constant at large distances from the galactic centre (purple line in Fig.~\ref{fig:DM_rotation_curve}), rather than follow the Kepplerian dependence $v \propto 1/\sqrt{r}$ (pink line in Fig.~\ref{fig:DM_rotation_curve}), which is expected from the observation that most stars are concentrated in the galactic core. These observations provide strong evidence for the presence of DM in galaxies, which is predominantly located at moderately large distances away from the galactic centre. DM is a non-luminous, non-baryonic form of matter than interacts very weakly with itself and Standard Model (SM) matter. Observations of stellar orbital velocities in our local galactic neighbourhood give the cold (non-relativistic) DM energy density within our local galactic neighbourhood of \cite{PDG2014}:
\begin{equation}
\label{rho_CDM_local}
\rho_{\textrm{CDM}}^{\textrm{local}} = 0.4~\textrm{GeV/cm}^3 .
\end{equation}
Further evidence for the existence of DM comes from gravitational lensing observations of the Bullet Cluster \cite{Bullet2003,Markevitch2006A,Markevitch2006B}, angular fluctuations in the cosmic microwave background (CMB) spectrum \cite{CMB2009}, and the need for non-baryonic matter to explain observed structure formation \cite{Silk2005Review}. The latest Wilkinson Microwave Anisotropy Probe (WMAP) observations give a present-day mean DM energy density of \cite{PDG2014}:
\begin{equation}
\label{rho_DM_mean-WMAP}
\bar{\rho}_{\textrm{DM}} = 1.3 \times 10^{-6}~\textrm{GeV/cm}^3 .
\end{equation}

\begin{figure}[h!]
\begin{center}
\includegraphics[width=4.5cm]{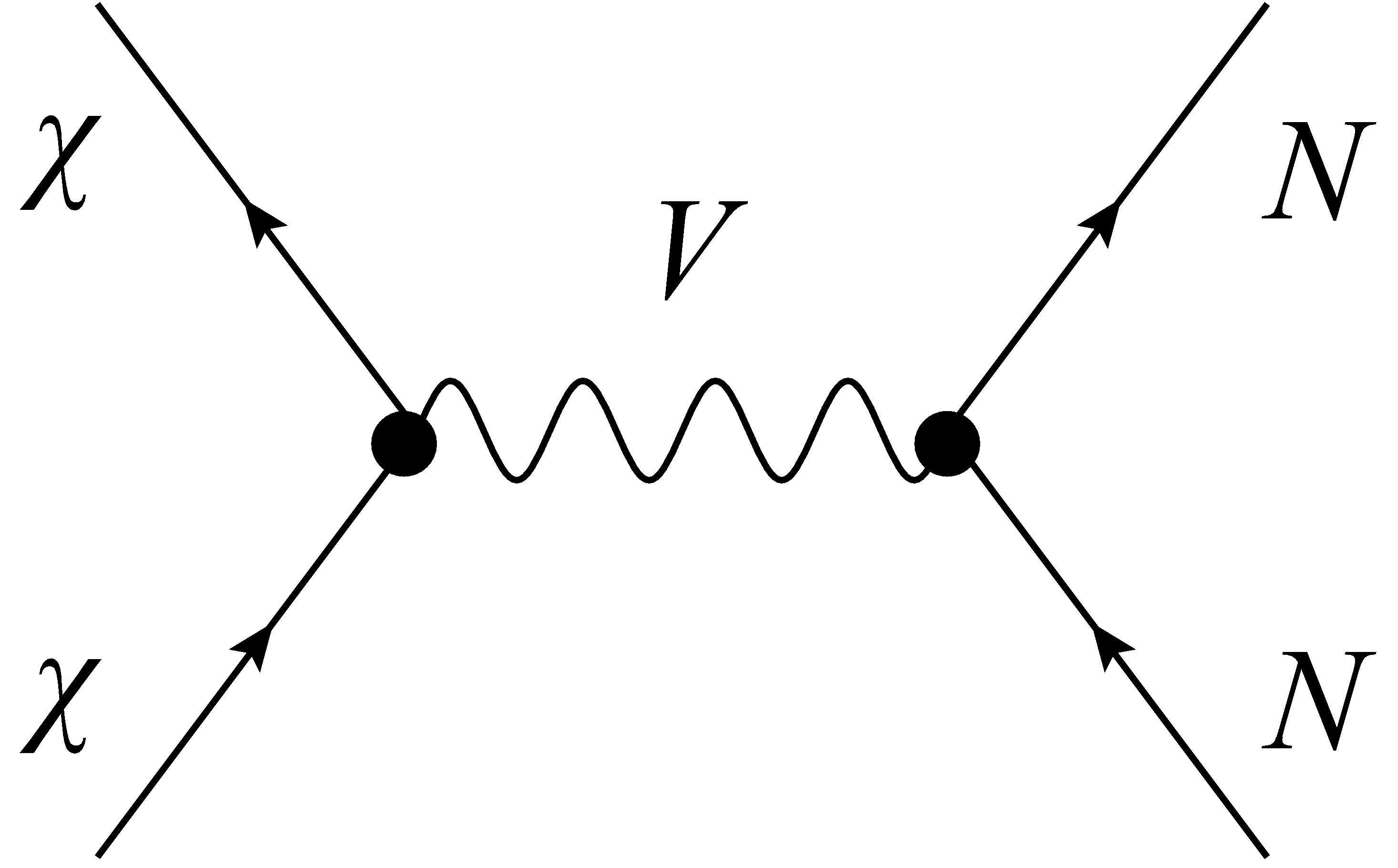}
\caption{Traditional WIMP dark matter detection experiments search for the scattering of WIMP dark matter off nuclei. The scattering cross-section associated with this process scales as: $\sigma_{\textrm{scat}} \propto (e'_\chi e'_N / M_V^2)^2$. } 
\label{fig:DM-N_scat}
\end{center}
\end{figure}

In order to explain its observed abundance, it is reasonable to expect that DM has non-gravitational interactions with ordinary matter. Despite the overwhelming evidence for its existence, direct searches for DM via non-gravitational interactions with ordinary matter have not yet produced a strong positive result, leaving the identity and non-gravitational interactions of DM in a state of mystery. Various theoretically well-motivated candidates for DM have been proposed, including weakly interacting massive particles (WIMPs), axions, and weakly interacting slim particles (WISPs). We refer the reader to the comprehensive reviews in Refs.~\cite{PDG2014,Bertone2010,Kim2010RMP-axions,Baer2015_DM-Review} for an overview of the theoretical motivation behind the main candidates, their role in cosmology and the main searches for these candidates. In traditional searches for WIMP DM (see e.g.~Refs.~\cite{SuperCDMS2014,CoGeNT2011,CRESST2014,DAMA2008,LUX2013,XENON2013}), which look for the scattering of WIMP DM off nuclei (Fig.~\ref{fig:DM-N_scat}), the sought effect is quartic in the underlying interaction parameters $e'$ that parametrise the interaction between DM and nucleons:
\begin{equation}
\label{DM-WIMP_4o}
\mathcal{L}_{\textrm{eff}} = \frac{e'_\chi e'_N}{4\pi} \frac{1}{ M_V^2} (\bar{\chi} \gamma^\mu \chi) (\bar{N} \gamma_\mu N) .
\end{equation}
The smallness of the interaction parameters $e'$ make further progress in these searches for WIMP DM very challenging. 

In recent times, there has been a growing interest to use atomic and related systems to directly search for DM. There is very strong motivation for the use of such systems, which to date have been employed with great success as high-precision frequency standards, in tests of the SM and as sensitive probes of new physics beyond the SM \cite{Khriplovich1991PNCBook,Ginges2004Review,Pospelov2005ReviewEDM,Kostelecky2011,Roberts2015Review}. 
Atomic clocks are one of the most precise instruments every built by mankind, with the best current fractional inaccuracies of the order $10^{-18}$ \cite{Hinkley2013YbClock,Ye2014SrClock,Katori2015Hg-Sr_Clock}. 
Experiments with atomic Hg provide the most precise limits on the electric dipole moment (EDM) of the proton, quark chromo-EDM and $P$,$T$-odd nuclear forces, as well as the most precise limits on the neutron EDM and Quantum Chromodynamics (QCD) $\theta$ term from atomic or molecular experiments \cite{Heckel2009HgEDM} (ultracold neutron experiments give the best limits for the latter parameters \cite{Baker2006NEDM}), while experiments with molecular ThO provide the most precise limit on the electron EDM \cite{DeMille2014ThO-ACME_EDM}. Measurements and calculations of the $6s$-$7s$ parity nonconserving (PNC) amplitude in atomic Cs stand as the most precise atomic test of the SM electroweak theory to date, see e.g.~Refs.~\cite{Bouchiat1982CsPNC,Wieman1997CsPNC,Dzuba1989CsPNC,Dzuba2012PNC}, and are competitive with direct searches performed at hadron colliders \cite{PDG2014,Abulencia2006}. Experiments with atomic co-magnetometers \cite{Berglund1995Hg-CsCPT,Walsworth2004He-XeCPT,Romalis2010He-KCPT,Peck2012Cs-HgCPT,Allmendinger2014Xe-HeCPT}, torsion pendula containing spin-polarised electrons \cite{Heckel2006eCPT,Heckel2008eCPT}, and ultracold neutrons \cite{Altarev2009nCPT} provide some of the most stringent limits on $CPT$- and Lorentz-invariance-violating physics. Laser interferometer experiments have set the most stringent limits on gravitational wave detection to date \cite{TAMA2004,LIGO2012,GEO2014}.

In the present book review, we present an overview of recent developments in the detection of light bosonic dark matter, including axion, pseudoscalar axion-like and scalar dark matter, which form either a coherently oscillating classical field or topological defects (solitons), using a variety of high-precision laboratory measurements. We particularly emphasise new measurements, in which the sought effects are linear in the underlying interaction strength between dark matter and ordinary matter, and are easier to search for than traditional quartic effects such as those shown in Fig.~\ref{fig:DM-N_scat}. New terrestrial experiments include measurements with atomic clocks, spectroscopy, atomic and solid-state magnetometry, torsion pendula, ultracold neutrons, and laser interferometry. Astrophysical observations, such as Big Bang nucleosynthesis (BBN) and CMB measurements, assist in terrestrial searches by ruling out regions of the relevant parameter spaces. New astrophysical observations that involve pulsar timing and cosmic radiation lensing can also be used to directly search for DM. We also discuss various recently proposed mechanisms for the induction of slow `drifts', oscillating variations and transient-in-time variations of the fundamental constants of Nature by DM, which offer a more natural means of producing a cosmological evolution of the fundamental constants compared with traditional dark energy-type theories, which invoke a (nearly) massless underlying field. 



\section{Axions}
\label{Axions++}
\subsection{The strong CP problem and the QCD axion}
\label{Strong_CP}
\begin{figure}[h!]
\begin{center}
\includegraphics[width=3.5cm]{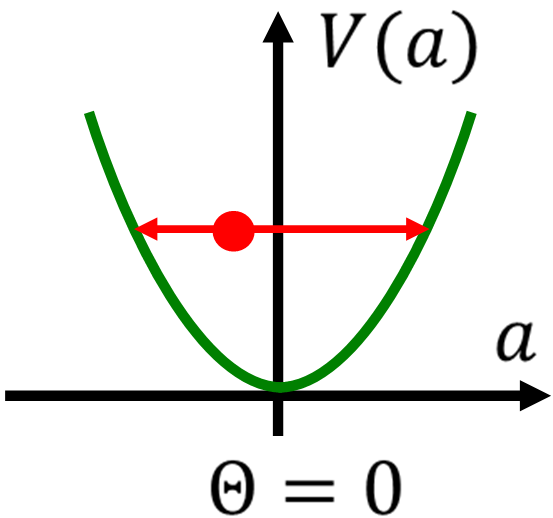}
\caption{The QCD axion oscillates about the minimum of its potential, which corresponds to $\theta=0$, thereby alleviating the strong $CP$ problem. The frequency of oscillation is set by the mass of the QCD axion $m_a$.} 
\label{fig:Axion_potential}
\end{center}
\end{figure}

When the SM was been developed during the 1970s, it quickly became apparent that there was an issue in the QCD sector as far the combined charge-parity ($CP$) symmetry was concerned. The QCD Lagrangian contains the $P$,$CP$-violating term \cite{Polyakov1975,Hooft1976,Jackiw1976,Gross1976}: 
\begin{equation}
\label{theta-term}
\mathcal{L}_{\textrm{QCD}}^{\theta} = \theta \frac{g^2}{32 \pi^2} G\tilde{G} ,
\end{equation}
where $\theta$ is the angle that quantifies the amount of $CP$ violation within the QCD sector, $g^2/4\pi = 14.5$ is the colour coupling constant, and $G$ and $\tilde{G}$ are the gluonic field tensor and its dual, respectively. Account of weak interaction effects results in a shift of $\theta$ from its bare value to the observable value $\bar{\theta}$ \cite{Peccei1981}. The angle $\bar{\theta}$ may in principle have assumed any value in the range $-\pi \le \bar{\theta} \le +\pi$, but its observed value from measurements of the permanent static neutron EDM is constrained to be $|\bar{\theta}| < 10^{-10}$ \cite{Baker2006NEDM}. The smallness of the observed value of $\bar{\theta}$ constitutes the strong $CP$ problem. An elegant and the most widely accepted resolution of the strong $CP$ problem was proposed by Peccei and Quinn \cite{PQ1977A,PQ1977B}, in which the $\theta$ parameter was interpretted as a dynamical field (the massive pseudoscalar axion, $a$):~$\bar{\theta} \to a(t)/f_a$, where $f_a$ is the axion decay constant. 
Initially, the axion field is constant ($\bar{\theta} \sim 1$ in the absence of fine-tuning of the vacuum misalignment angle $\theta_1$), but for times when $m_a \gg H$, where $H$ is the Hubble constant, the axion undergoes oscillations about the minimum of its potential (Fig.~\ref{fig:Axion_potential}), which corresponds to $\bar{\theta}=0$, hence alleviating the strong $CP$ problem \cite{Wilczek1983,Abbott1983,Dine1983}.

\subsection{Axions as cold dark matter}
\label{Axions_CDM}

Although the original PQWW model of the axion \cite{PQ1977A,PQ1977B,Weinberg1978,Wilczek1978} was quickly ruled out experimentally, the KSVZ \cite{Kim1979,Shifman1980} and DFSZ \cite{Zhitnitsky1980,Dine1981} models of the QCD axion turned out to be compatible with all terrrestrial and astrophysical observations (for some of the more recent invisible axion models based on the Peccei-Quinn symmetry, we refer the reader to Refs.~\cite{Lindner2010PQ,Celis2014PQ,Celis2015PQ,Anh2015PQ}). The properties of the QCD axion are predominantly determined by the axion decay constant $f_a$. In particular, the QCD axion mass $m_a$ is related to $f_a$ via the relation
\begin{equation}
\label{QCD_axion_mass}
m_a \sim 6 \times 10^{-5}~\textrm{eV} \left( \frac{10^{11}~\textrm{GeV}}{f_a} \right) .
\end{equation}
For very weak couplings (i.e.~for very large values of $f_a$), axions are produced non-thermally in the early Universe. At temperatures well above the QCD phase transition, the QCD axion is effectively massless and the corresponding field can take any value, parameterised by $\theta_1$. The axion develops its non-zero mass $m_a$ (due to nonperturbative effects) when the temperature $T \lesssim \textrm{GeV}$, and for times when $m_a \gg H$, the axion undergoes oscillations about the minimum of its potential at $a = 0$ (Fig.~\ref{fig:Axion_potential}). The resulting axion energy density, produced via this vacuum misalignment mechanism, is given (in terms of the critical energy density) by \cite{Wilczek1983,Abbott1983,Dine1983}:
\begin{align}
\label{VMM_Axion_density}
\Omega_{\textrm{axion}} \sim \theta_1^2 \left( \frac{f_a}{10^{12} ~ \textrm{GeV}} \right)^{1.18} .
\end{align}
For $\theta_1 \sim 1$, axions saturate the present-day CDM content if $f_a \sim 10^{12}$ GeV. For $f_a \gg 10^{12}$ GeV, axion production via the vacuum misalignment mechanism would have led to the overclosure of the Universe unless $\theta_1 \ll 1$, which may arise due to fine-tuning of the vacuum misalignment angle or anthropic selection \cite{Weinberg1987,Linde1988,Wilczek2008}. We also note that the density of axions produced in the early Universe depends on the order of the cosmological events, in particular whether the Peccei-Quinn symmetry is broken prior to or following cosmic inflation. In the latter case, there may be additional contributions to the axion density of the same order as in Eq.~(\ref{VMM_Axion_density}) from the formation and decay of axionic topological defects, such as cosmic strings and domain walls \cite{Sikivie2008LN}.


Axions produced by the vacuum misalignment mechanism are very cold with almost no kinetic energy. Furthermore, if they are sufficiently light and weakly interacting, then these axions may survive until the present day and reside in the observed galactic DM haloes. If $m_a < 2 m_e$, then the axion lifetime is determined by its two-photon decay channel \cite{Kim2010RMP-axions}:
\begin{align}
\label{a-gamma_lifetime}
\tau (a \to 2 \gamma) = \frac{2^8 \pi^3}{C_\gamma^2 \alpha^2} \frac{f_a^2}{m_a^3} ,
\end{align}
where $|C_\gamma| \sim 1$ is a model-dependent coefficient. For $m_a \lesssim 24$ eV, the axion lifetime exceeds the present age of the Universe. Thus, ultralight (sub-eV mass) axions, as well as axion-like pseudoscalar particles (ALPs) and scalar particles, for which no predictive mass formula akin to Eq.~(\ref{QCD_axion_mass}) exists, are good candidates for cold DM. Ultralight spin-0 bosons are good candidates for the dominant contributor to cold DM to very low particle masses. The simplest model-independent lower limit arises from the requirement that the de Broglie wavelength of the DM particles not exceed the halo size of the smallest galaxies, giving $m_a \gtrsim 10^{-22}$ eV. This simple estimate is in fact in good agreement with more rigorous limits obtained from cosmological and astrophysical investigations. Ultralight spin-0 DM would have inhibited cosmological structure growth \cite{Marsh2013DMHalo} in conflict with Lyman-alpha observations \cite{Viel2013warmDM}, unless $m_a \gtrsim 10^{-21}$ eV. Ultralight spin-0 DM would have suppressed high-redshift galaxy formation, contrary to observations unless $m_a \gtrsim 10^{-22}$ eV \cite{Marsh2015ULA_reionise}, while CMB observations necessitate $m_a \gtrsim 10^{-24}$ eV \cite{Marsh2015ULA_cosmo}. We stress that these constraints only apply for ultralight spin-0 DM which is the dominant contributor to cold DM. Due to its effects on structure formation, ultralight spin-0 DM in the mass range $10^{-24} - 10^{-20}$ eV has been proposed \cite{Gruzinov2000fuzzyDM,Marsh2013DMHalo,Schive2014A,Schive2014B} to solve several long-standing astrophysical puzzles, such as the cusp-core, missing satellite, and too-big-to-fail problems \cite{Weinberg2015ReviewDMIssues} (see also the earlier work of Ref.~\cite{Khlopov1985scalar}). 

We note in passing that, while most interest resides in cold, non-relativistic, ultralight spin-0 bosons, the possibility of relativistic spin-0 bosons in the early Universe has also also been investigated. Relativistic particle species increase the total energy density in the early Universe ($\rho_{\textrm{rel}} \propto (1+z)^4$) and in turn increase the rate of cosmic expansion, which is parametrised by $H$ (non-relativistic species, for which $\rho_{\textrm{non-rel}} \propto (1+z)^3$, do not affect the total energy density appreciably at early times). The presence of additional beyond-the-SM relativistic particle species, therefore, increases the neutron-to-proton ratio at the time of weak interaction freeze-out, $n/p = e^{-(m_n - m_p)/T_F}$, by causing freeze-out of the weak interactions, $p + e^- \leftrightarrow n + \nu_e$ and the corresponding crossing reactions, to occur at an earlier time (which corresponds to a larger freeze-out temperature $T_F$). Measurements and SM calculations of the primordial $^4$He abundance allow for an additional, relativistic, neutral spin-0 particle during BBN, while an additional relativistic spin-$1/2$ or spin-1 particle is excluded (the combination of the $^4$He abundance and the CMB value for the baryon-to-photon ratio $\eta$ do not alter this conclusion) \cite{Olive2005}.

The number density of ultralight spin-0 fields per de Broglie volume readily exceeds unity, $n_a/\lambda_{\textrm{dB}}^3 \gg 1$. As a result, these bosons readily form a coherently oscillating classical field
\begin{equation}
\label{classical_field_axion_no_KE}
a(t) \simeq a_0 \cos(m_a t) ,
\end{equation}
with an amplitude $a_0 \simeq \sqrt{2 \rho_{\textrm{axion}}}/ m_a$, where $\rho_{\textrm{axion}}$ is the energy density associated with the bosonic field, and with a very well-defined oscillation frequency set by the boson mass.
Over time, gravitational collapse of ultralight bosons into galaxies and their interaction with ordinary matter resulted in their virialisation, which led to a loss of perfect monochromaticity in their oscillation frequency
\begin{equation}
\frac{\Delta \omega_a}{m_a} \sim v_{\textrm{virial}}^2 \sim 10^{-6} ,
\end{equation}
where a virial velocity $v_{\textrm{virial}} \sim 10^{-3}$ is typical within our local galactic region. In the moving reference frame of our Solar System, a bosonic field has a non-zero average momentum $\v{p}_a$ and so the bosonic field takes the form
\begin{equation}
\label{classical_field_axion_virialised}
a(\v{r},t) = a_0 \cos( \omega_a t - \v{p}_a \cdot \v{r} ) .
\end{equation}
Despite the loss of perfect monochromaticity, the bosonic field remains coherent on time scales less than the coherence time
\begin{equation}
\tau_{\textrm{coh}} \sim \frac{2\pi}{m_\phi v_{\textrm{virial}}^2} \sim 10^6 \left(\frac{2\pi}{m_\phi} \right) ,
\end{equation}
which is determined by the criterion that the additional phase accumulated over time in Eq.~(\ref{classical_field_axion_virialised}) due to virialisation remains less than $2\pi$.

\subsection{Axion interactions and astrophysical constraints}
\label{Axions+Astrophys}
The axion couples to SM particles as follows (we consider only the couplings that are of direct interest to experimental searches):
\begin{align}
\label{Axion_couplings}
\mathcal{L}_{\textrm{axion}} = \frac{a}{f_a} \frac{g^2}{32\pi^2} G\tilde{G} + \frac{C_\gamma a}{f_a} \frac{e^2}{32\pi^2} F\tilde{F} - \sum_{f} \frac{C_f}{2f_a} \partial_\mu a ~ \bar{f} \gamma^\mu \gamma^5 f ,
\end{align}
where the first term represents the coupling of the axion field to the gluonic field tensor $G$ and its dual $\tilde{G}$, the second term represents the coupling of the axion field to the electromagnetic field tensor $F$ and its dual $\tilde{F}$, while the third term represents the coupling of the derivative of the axion field to the fermion axial-vector currents $\bar{f} \gamma^\mu \gamma^5 f$. $C_\gamma$ and $C_f$ are model-dependent coefficients. Typically, $|C_\gamma| \sim 1$ and $|C_n| \sim |C_p| \sim 1$ in models of the QCD axion \cite{Kim2010RMP-axions,Srednicki1985axion}. Within the DFSZ model, where the tree level coupling of the axion to the electron is non-vanishing, $|C_e| \sim 1$ \cite{Srednicki1985axion}. However, within the KSVZ model, $|C_e| \sim 10^{-3}$, since the tree level coupling vanishes and the dominant effect arises at the 1-loop level \cite{Srednicki1985axion}. For ALPs, the coefficients $C_\gamma$ and $C_f$ are essentially free parameters, and the coupling to gluons is generally presumed absent. The common parameter in Eq.~(\ref{Axion_couplings}), which is of the most interest, is the axion decay constant $f_a$.

Astrophysical constraints on axion parameters greatly assist in laboratory searches for axions. For stellar axions, consideration of the axion production processes $\gamma + \gamma \to a$, $\gamma + e^- \to a + e^-$, $\gamma + N \to a + N$, the axion absorption processes $a + e^- \to \gamma + e^-$, $a + N \to \gamma + N$ and the decay channel $a \to \gamma + \gamma$ in stars gives the following lower bound on the mass of stellar axions \cite{Khlopov1978}:
\begin{equation}
\label{solar_axion_mass-limit}
m_a^{\textrm{stellar}} > 25~\textrm{keV} ,
\end{equation}
while the requirement that stellar axion emission from helium-burning red giants not disrupt observed stellar evolution gives the stronger lower limit of \cite{Khlopov1978,Dicus1978}:
\begin{equation}
\label{Red_giant_axion_mass-limit}
m_a^{\textrm{stellar}} \gtrsim 200~\textrm{keV} .
\end{equation}
Energy loss from stars by solar axion emission ($\gamma + \gamma \to a$) requires enhanced nuclear burning, which would lead to an increase in the solar $^{8}$B neutrino flux that contradicts observations unless the axion couples sufficiently weakly to the photon \cite{Raffelt2008LN}:
\begin{equation}
\label{C_gamma_limit_Sun}
f_a / C_\gamma \gtrsim 2 \times 10^{6} ~\textrm{GeV} .
\end{equation}
The application of energy-loss arguments to the nucleon bremmstrahlung process $N + N \to N + N + a$ in supernovae 1987A gives the following limit on the axion coupling to nucleons \cite{Raffelt2008LN}:
\begin{equation}
\label{C_N_limit_SN1987A}
f_a / C_N \gtrsim 10^{9} ~\textrm{GeV} ,
\end{equation}
Energy loss mechanisms in stars through the Compton-like process $\gamma + e^- \to a + e^-$ and through the bremsstrahlung process $e^- + (Z,A) \to  e^- + (Z,A) + a$ would excessively delay the onset of helium burning unless the axion couples sufficiently weakly to the electron \cite{Raffelt2008LN}:
\begin{equation}
\label{C_e_limit_WD_cooling}
f_a / C_e \gtrsim 2 \times 10^{9} ~ \textrm{GeV} ,
\end{equation}
with similar constraints from consideration of the increase in white dwarf cooling rates due to axion emission \cite{Raffelt2008LN,Isern2001}.


\subsection{Traditional axion searches}
\label{Axions+tradn_searches}

\begin{figure}[h!]
\begin{center}
\includegraphics[width=4cm]{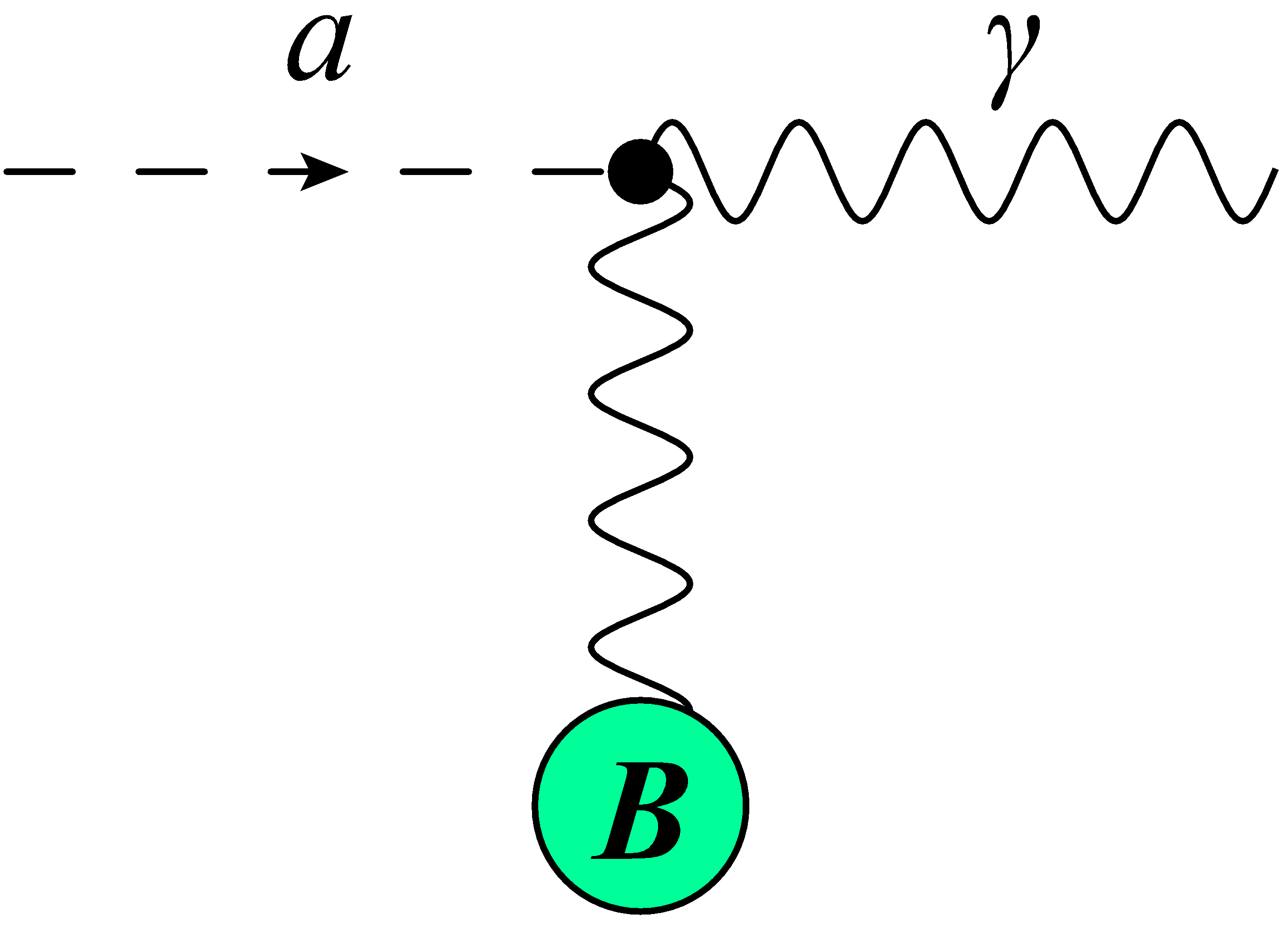}
\caption{Haloscope and helioscope experiments search for the conversion of galactic and solar axions, respectively, into photons in a strong applied magnetic field. The expected power generated by the conversion $a \to \gamma$ scales as: $P_{a \to \gamma} \propto (1/f_a)^2$.} 
\label{fig:Axion-photon_vertex}
\end{center}
\end{figure}

\begin{figure}[h!]
\begin{center}
\includegraphics[width=6.2cm]{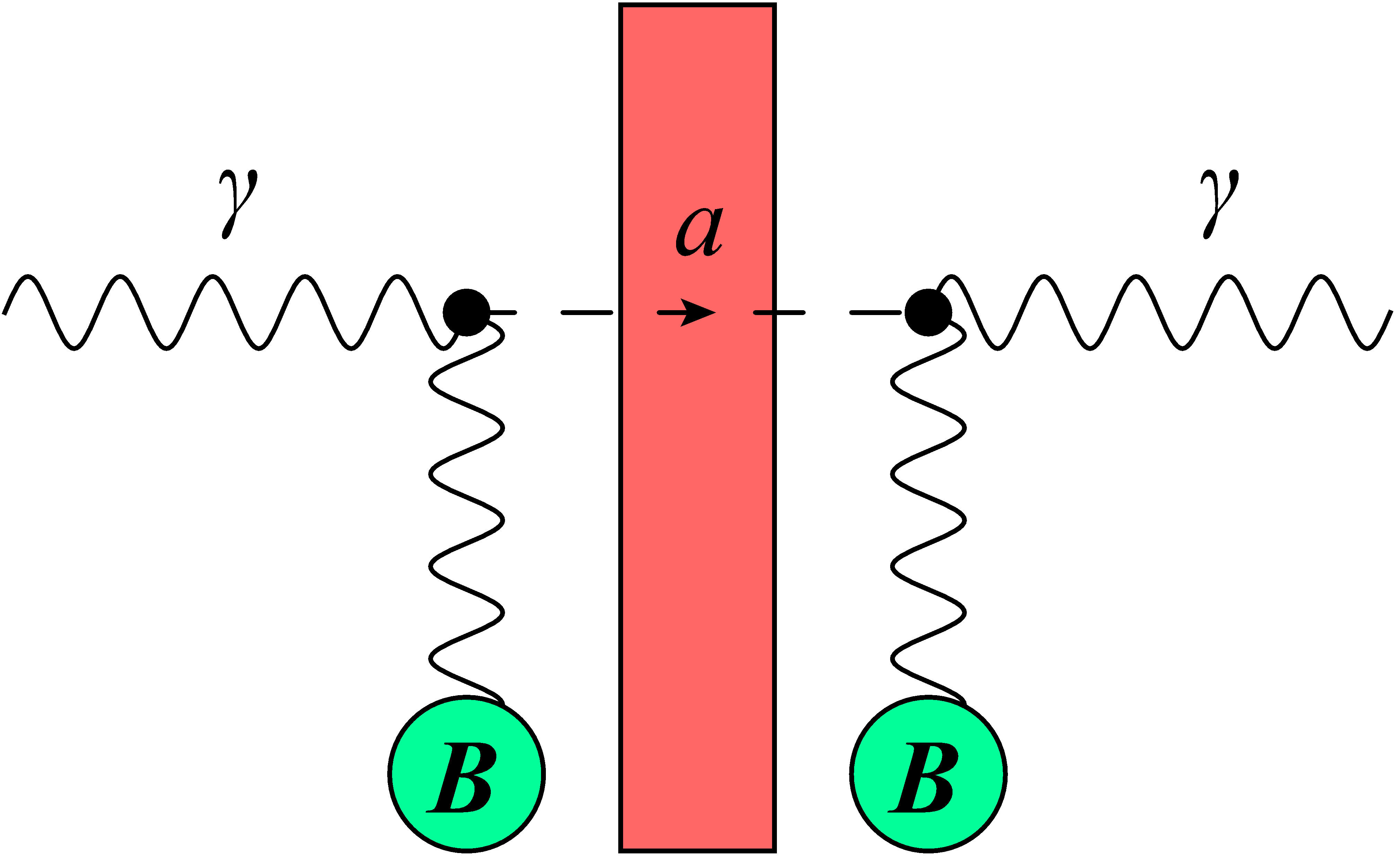}
\caption{`Light-shining-through-wall' experiments search for the transmission of photons through an impermeable material, due to the interconversion $\gamma \to a \to \gamma$ in the presence of strong applied magnetic fields on either side of the barrier. The expected power generated by the interconversion $\gamma \to a \to \gamma$ scales as: $P_{\gamma \to a \to \gamma} \propto (1/f_a)^4$.} 
\label{fig:Axion-photon_vertex_sqrd}
\end{center}
\end{figure}

Haloscope and helioscope methods can be used to search for galactic and solar axions, respectively, via the axion's coupling to the photon \cite{Sikivie1983ADMX}. The traditional haloscope (ADMX) \cite{ADMX2010} and helioscope (CAST) \cite{CAST2014} experiments (Fig.~\ref{fig:Axion-photon_vertex}) have shed valuable light on our understanding on the possible axion parameter space for the axion-photon coupling. The IAXO helioscope experiment \cite{IAXO2014} will be the upgrade of the present CAST experiment. Searches for solar axions via the axio-electric effect with scintillator detectors have also been conducted \cite{Avignone1987,Derbin2009,Derbin2012}. Various `light-shining-through-wall' \cite{GammeV2008,ALPS2013,CROWS2013} (Fig.~\ref{fig:Axion-photon_vertex_sqrd}) and vacuum birefringence \cite{BMV2007,PVLAS2013} searches for axions and ALPs via the axion-photon coupling have also been performed (for an overview of light-shining-through-wall searches for ALPs and various other light bosonic DM particles, we refer the reader to the review \cite{LSW_Review_2011}).

\begin{figure}[h!]
\begin{center}
\includegraphics[width=4.7cm]{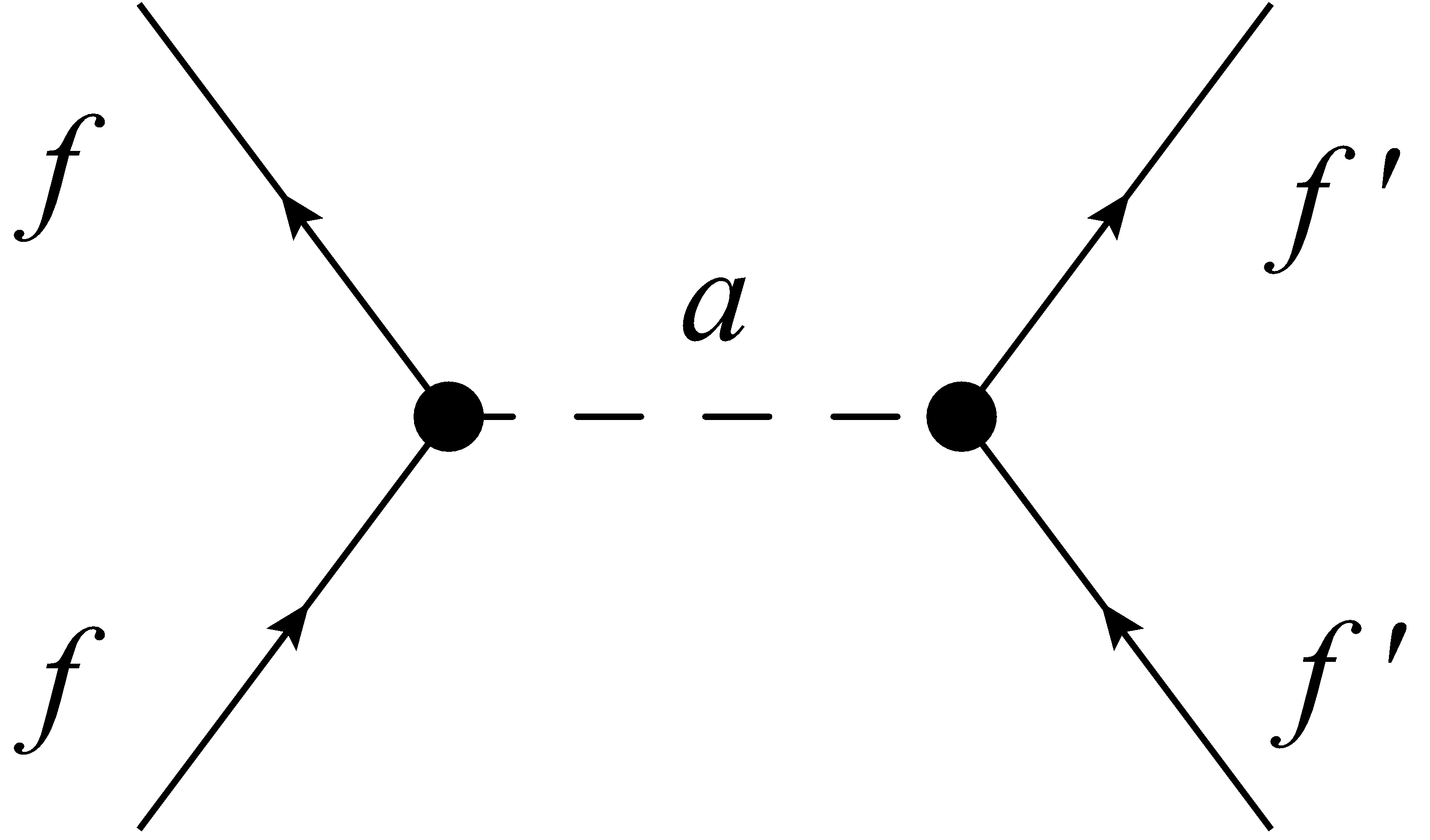}
\caption{Tabletop experiments can search for the effects of new macroscopic forces mediated by the exchange of ultralight spin-0 bosons. The induced energy shift due to such a new force scales as: $\delta \varepsilon \propto (1/f_a)^2$. Two vertices of scalar form $g_s \phi \bar{\psi} \psi$ produce a spin-independent monopole-monopole force. A vertex of scalar form, combined with a vertex of pseudoscalar form $g_p \phi \bar{\psi} i \gamma_5 \psi$, produce a $P,T$-violating monopole-dipole force. Two vertices of pseudoscalar form produce a fully spin-dependent dipole-dipole force.} 
\label{fig:New_force_ALP}
\end{center}
\end{figure}

Searches for ultra-light spin-0 bosons in tabletop experiments via the macroscopic forces they would produce due to their couplings with the electron and nucleons (Fig.~\ref{fig:New_force_ALP}) have also been proposed \cite{Wilczek1983_New-force}. Exchange of spin-0 bosons between fermions can produce either a spin-independent monopole-monopole potential, a $P,T$-violating monopole-dipole potential with the $\v{\sigma} \cdot \hat{\v{r}}$ correlation, or a fully spin-dependent dipole-dipole potential. Atomic magnetometry \cite{Lamoreaux1996_NF,Walsworth2008_NF,Romalis2009_NF,Petukhov2010_NF,Tullney2013_NF,Chu2013_NF,Bulatowicz2013_NF,Kimball2015_NF}, torsion pendulum \cite{Heckel2011_NF}, differential force measurements \cite{Mostepanenko2015_NF} and ultracold neutron experiments \cite{Baessler2007_NF,Serebrov2010_NF,Grenoble2015_NF} have collectively probed the axion-electron and axion-nucleon couplings over an expansive range of axion masses (see also Ref.~\cite{Hunter2013geoelectrons} for constraints on long-range interactions between spin-polarised geoelectrons deep within the Earth and the spin-polarised electrons and nucleons in laboratory experiments, and Ref.~\cite{Raffelt2012} for constraints on axion-electron and axion-nucleon interactions from a combination of terrestrial equivalence principle tests and astrophysical energy-loss bounds). Constraints on axion interactions through their mediation of spin-spin couplings in atomic systems have also been derived \cite{Karshenboim2011axion}.

\subsection{New axion searches}
\label{Axions+new_searches}
A number of new proposals to search for axions have been put forward over the recent years. These proposals may be partitioned into two broad categories: 

(I) New-generation searches, where the underlying axion-induced effects are \emph{linear} in the combination $a/f_a$ ($a_0/f_a \simeq 4 \times 10^{-19}$ for the QCD axion, which obeys Eq.~(\ref{QCD_axion_mass}) and which also saturates the local cold DM energy density in Eq.~(\ref{rho_CDM_local})), and are intrinsically much larger than the effects in traditional searches of Sec.~(\ref{Axions+tradn_searches}) and those in (II) below. These new-generation searches are outlined in Secs.~(\ref{Sec:Axion_wind}) $-$ (\ref{Sec:LC_Circuit}) that follow.

(II) Searches, where the underlying effects are quadratic or higher order in the combination of axion parameters $a/f_a$. These proposals are summarised in Sec.~(\ref{Sec:other}) below.

\subsubsection{`Axion wind' effect}
\label{Sec:Axion_wind}
The $\mu=1,2,3$ components in the coupling of the derivative of the axion field to the fermion axial-vector currents in Eq.~(\ref{Axion_couplings}) lead to the following non-relativistic Hamiltonian \cite{Flambaum2013Patras,Stadnik2014axions,Graham2013}:
\begin{equation}
\label{axion_wind}
H_{\textrm{eff}}(t) = \sum_{f=e,p,n} \frac{C_f a_0}{2f_a} \sin(m_a t) ~ \v{p}_a \cdot \v{\sigma}_{f} ,  
\end{equation}
which implies that a spin-polarised source of particles interacts with the axion 3-momentum, producing oscillating shifts in the energy of the spin-polarised source (which are linear in $a_0/f_a$) at two characteristic frequencies: $\omega_1 \simeq m_a$ and $\omega_2 = 2\pi/T_{\textrm{sidereal}}$, where $T_{\textrm{sidereal}} = 23.93$ hours is the sidereal day duration. This is the `axion wind' effect, which may be sought for using a variety of spin-polarised sources, for example, atomic co-magnetometers, torsion pendula and ultracold neutrons. 

Distortion of the axion field by the gravitational field of a massive body, such as the Sun or Earth, results in an additional axion-induced oscillating spin-gravity coupling (oscillating gravi-magnetic moments): $H'_{\textrm{eff}} (t) \propto (C_f a_0/f_a) \sin(m_a t) ~ \v{\sigma}_f \cdot \v{\hat{r}}$, which is directed towards the centre of the gravitating body \cite{Stadnik2014axions}. For couplings of the axion to nucleons inside the nucleus, isotopic dependence ($C_n \ne C_p$) requires knowledge of the proton and neutron spin contributions in experiments that search for the `axion wind' effect and oscillating gravi-magnetic moments. The proton and neutron spin contributions for nuclei of experimental interest have been calculated in Ref.~\cite{Stadnik2015NMBE}.

\subsubsection{Transient `axion wind' effect}

Apart from the classical fields that ultralight axions and other spin-0 fields may form (see Sec.~(\ref{Axions_CDM})), ultralight bosonic DM fields may also form topological defects, which arise from the stabilisation of the DM field under a suitable self-potential \cite{tHooft1974,Polyakov1974,Kibble1976,Khlopov1978TD,Zeldovich1980,Vilenkin1982,Sikivie1982}. Topological defects, which make up a sub-dominant fraction of DM, are believed to function as seeds for structure formation \cite{Brandenberger1987}. For some of the more recent developments on topological defects, we refer the reader to Refs.~\cite{Kirk2014,Sugiyama2015,Harko2015,Marsh2015,Oksanen2015}, while for the classical review, we refer the reader to Ref.~\cite{Vilenkin1985REVIEW}.

While stable domain wall structures that consist of the QCD axion would lead to disastrous consequences in cosmology by storing too much energy \cite{Sikivie1982}, domain walls and other topological defect structures consisting of ALPs or scalars are viable for certain combinations of parameters. Topological defects that consist of ALPs may interact with fermion axial-vector currents via the $\mu=1,2,3$ components of the derivative coupling in Eq.~(\ref{Axion_couplings}), which in the non-relativistic limit reads \cite{GNOME2013A}:
\begin{equation}
\label{axion_wind_transient}
H_{\textrm{eff}}(t) = \sum_{f=e,p,n} \frac{C_f}{2f_a} \left( \v{\nabla} a \right) \cdot \v{\sigma}_{f} .
\end{equation}
Eq.~(\ref{axion_wind_transient}) implies that a spin-polarised source of particles will temporarily interact with a topological defect as the defect passes through the system.  A global network of detectors, such as atomic co-magnetometers, has been proposed to detect these correlated transient-in-time signals, produced by the passage of a defect through Earth \cite{GNOME2013A,GNOME2013B}.

\subsubsection{Oscillating $P$,$T$-odd electromagnetic moments}
\begin{figure}[h!]
\begin{center}
\includegraphics[width=4.4cm]{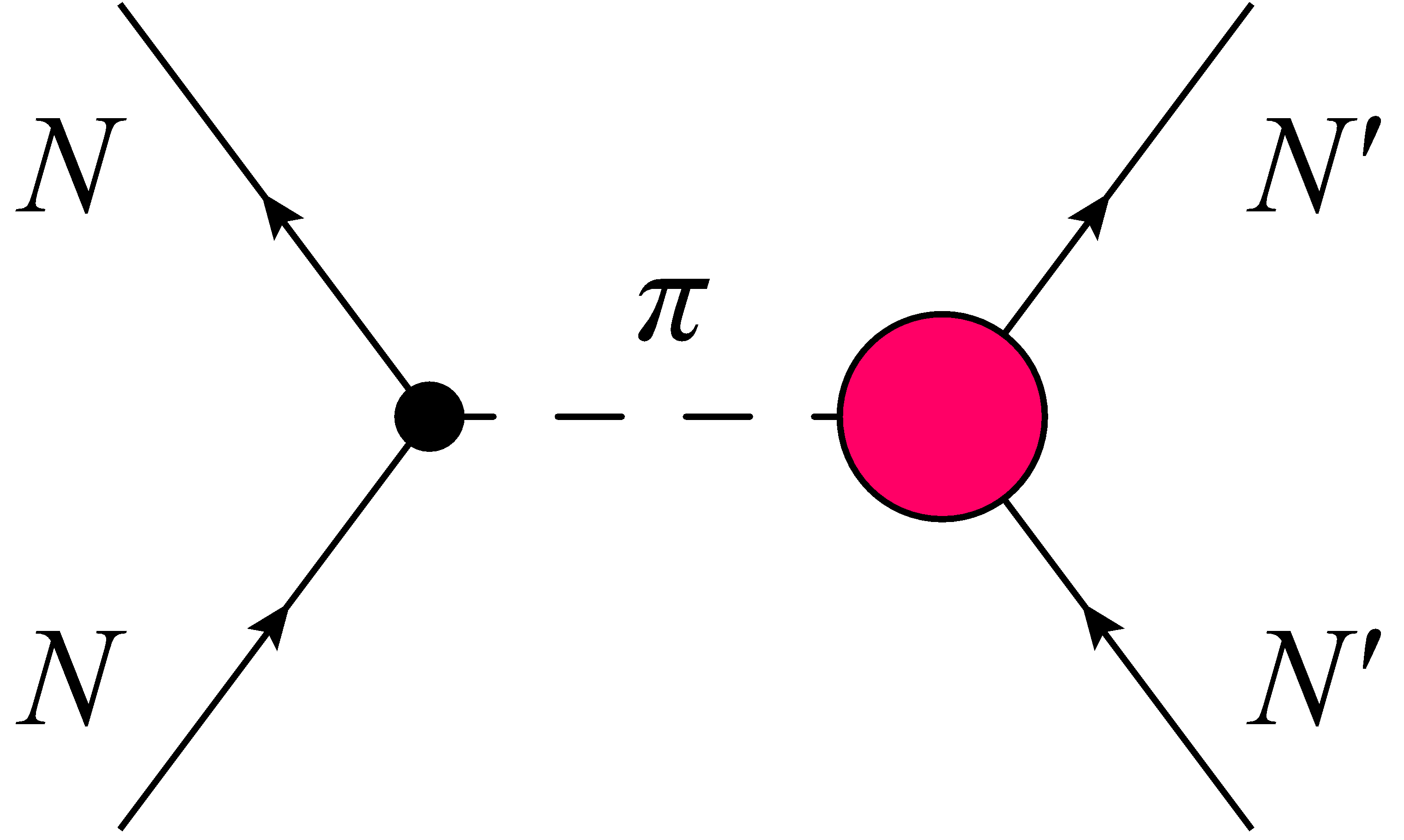}
\caption{Main process responsible for the induction of oscillating $P$,$T$-odd nuclear electromagnetic moments by an oscillating axion field. The black vertex on the left is due to the usual strong $P$,$T$-conserving $\pi N N$ coupling ($g_{\pi N N} = 13.5$), while the magenta vertex on the right is due to the axion-induced $P$,$T$-violating $\pi N N$ coupling ($\bar{g}_{\pi N N} \simeq 0.027 a_0/f_a \cos(m_a t)  $) \cite{Stadnik2014axions}. } 
\label{fig:Pion_nucleon_PT-odd}
\end{center}
\end{figure}

 Interaction of the QCD axion field with the gluon fields in Eq.~(\ref{Axion_couplings}) produces an oscillating neutron EDM \cite{Graham2011,Stadnik2014axions}:
\begin{equation}
\label{Osc_NEDM}
d_n (t) \simeq 1.2 \times 10^{-16} ~ \frac{a_0 }{f_a} \cos(m_a t) ~ e \cdot \textrm{cm} ,
\end{equation}
which induces oscillating nuclear Schiff moments \cite{Graham2011,Stadnik2014axions} and oscillating nuclear magnetic quadrupole moments \cite{Roberts2014long}. In nuclei, a second and more efficient mechanism exists for the induction of oscillating electromagnetic moments by axions --- namely, the \emph{P},\emph{T}-violating nucleon-nucleon interaction that is mediated by pion exchange, with the axion field supplying the oscillating source of \emph{P} and \emph{T} violation at one of the $\pi N N$ vertices \cite{Stadnik2014axions} (Fig.~\ref{fig:Pion_nucleon_PT-odd}).

Axion-induced oscillating $P$,$T$-odd nuclear electromagnetic moments can in turn induce oscillating EDMs in atoms and molecules. In diamagnetic species ($J=0$), only oscillating nuclear Schiff moments (which require $I \ge 1/2$) produce an oscillating atomic/molecular EDM (oscillating nuclear EDMs are effectively screened for typical axion masses, as a consequence of Schiff's theorem \cite{Schiff1963}). Two atoms that are of particular experimental interest are $^{199}$Hg and $^{225}$Ra, for which the axion induces the following oscillating EDMs \cite{Stadnik2014axions}:
\begin{align}
\label{AtomicEDM_axion-Hg}
d(^{199}\textrm{Hg}) = -1.8 \times 10^{-19} ~\frac{a_0}{f_a} \cos(m_a t) ~ e \cdot \textrm{cm} , \\
d(^{225}\textrm{Ra}) = 9.3 \times 10^{-17} ~\frac{a_0}{f_a} \cos(m_a t) ~ e \cdot \textrm{cm} ,
\end{align}
with the large enhancement in $^{225}$Ra compared with $^{199}$Hg due to both collective effects and small energy separation between members of the relevant parity doublet, which occurs in nuclei with octupolar deformation and results in a significant enhancement of the nuclear Schiff moment \cite{Auerbach1996,Auerbach1997}. A possible platform to search for the oscillating EDMs of diamagnetic atoms in ferroelectric solid-state media has been proposed in Ref.~\cite{CASPER2014}.

Paramagnetic species ($J \ge 1/2$) offer more rich possibilities. Firstly, axion-induced oscillating nuclear magnetic quadrupole moments (which require $I \ge 1$) also produce an oscillating atomic/molecular EDM \cite{Roberts2014long}, which is typically larger than that due to an oscillating nuclear Schiff moment (since magnetic quadrupole moments are not subject to screening of the applied electric field by atomic/molecular electrons). Secondly, an entirely different mechanism exists for the induction of oscillating EDMs in paramagnetic species, through the derivative interaction of the axion field with atomic/molecular electrons in Eq.~(\ref{Axion_couplings}). The $\mu = 0$ component of this term mixes atomic/molecular states of opposite parity (with both imaginary and real coefficients of admixture), generating the following oscillating atomic EDM (due to the real coefficients of admixture) in the non-relativistic approximation for an $S_{1/2}$ state \cite{Stadnik2014axions}:
\begin{equation}
\label{atomic_EDM_para}
d_a (t) \sim - \frac{C_e a_0 m_a^2 \alpha_s}{f_a \alpha} e \sin(m_a t) ,
\end{equation}
where $\alpha_s$ is the static scalar polarisability. Fully relativistic Hartree-Fock atomic calculations are in excellent agreement with the scaling $d_a \propto \alpha_s$ in Eq.~(\ref{atomic_EDM_para}) \cite{Roberts2014long,Roberts2014prl}. The imaginary coefficients of admixture in the perturbed atomic wavefunction produce \emph{P}-violating, \emph{T}-conserving effects in atoms, while the analogous imaginary coefficients of admixture in the perturbed nuclear wavefunction (due to the axion-nucleon interaction via the $\mu = 0$ component of the third term in Eq.~(\ref{Axion_couplings})) produce \emph{P}-violating, \emph{T}-conserving nuclear anapole moments \cite{Stadnik2014axions,Roberts2014long,Roberts2014prl}.

\begin{figure}[h!]
\begin{center}
\includegraphics[width=6.2cm]{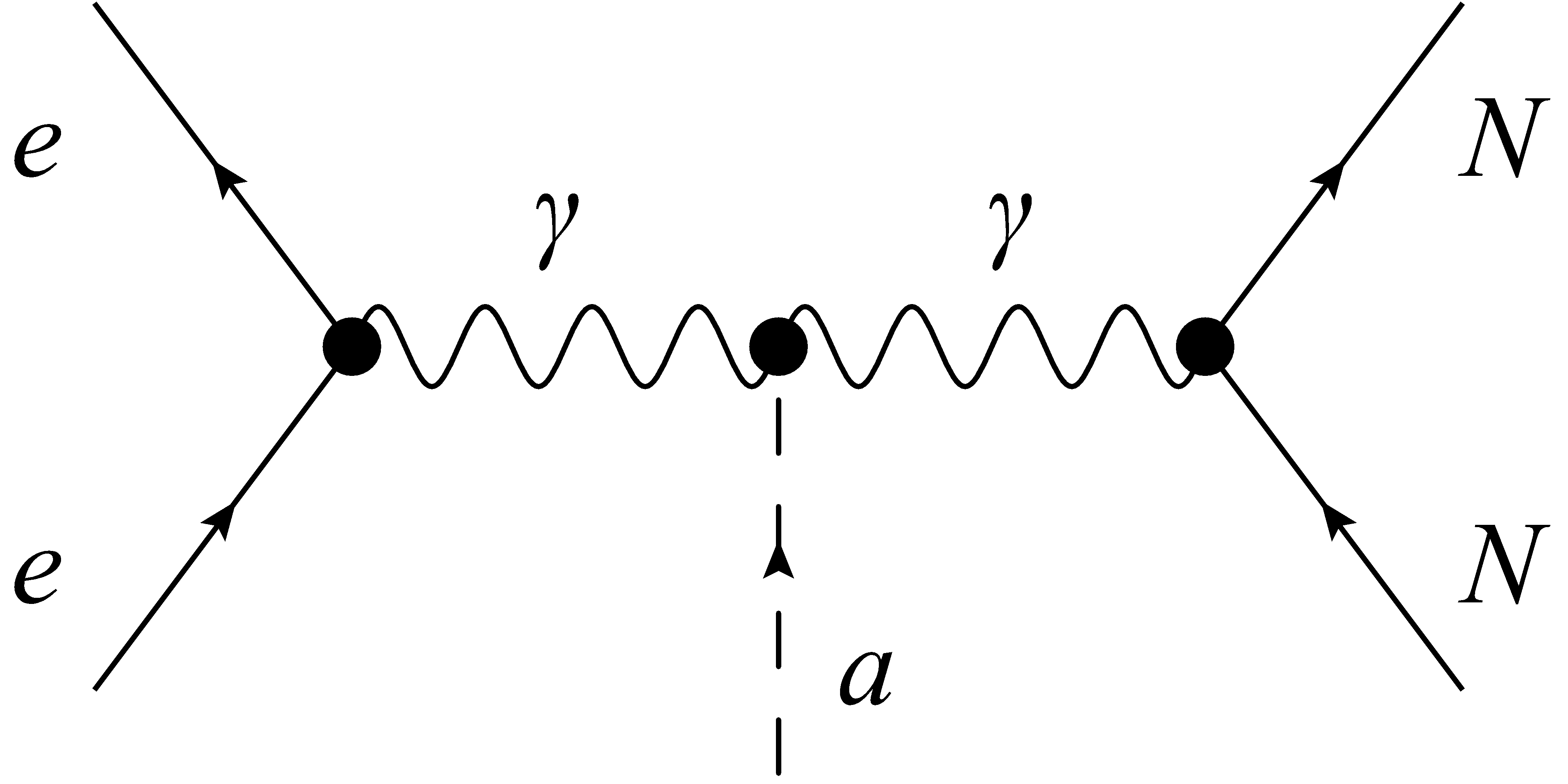}
\caption{Induction of an oscillating electric dipole moment in a paramagnetic atom or molecule may occur as a result of the perturbation of the Coulomb interaction of atomic/molecular electrons and nucleons by the axion electromagnetic anomaly. } 
\label{fig:Axion-electron-nucleon_anomalous_2}
\end{center}
\end{figure}

An axion or ALP field may also induce oscillating EDMs in paramagnetic species via perturbation of the electron-nucleon Coulomb interaction by the axion-photon interaction of Eq.~(\ref{Axion_couplings}) \cite{Flambaum2015axion} (Fig.~\ref{fig:Axion-electron-nucleon_anomalous_2}).

\subsubsection{Oscillating electric current flows along magnetic field lines} 
\label{Sec:LC_Circuit}
Interaction of an axion field with the photon field in Eq.~(\ref{Axion_couplings}) causes an oscillating electric current to flow along magnetic field lines \cite{Sikivie2014LC}. Conversely, in the presence of an externally applied magnetic field $\v{B}_0$, an axion field induces an electric current density 
\begin{equation}
\label{osc_current}
\v{j}_a = \frac{\alpha C_\gamma a_0  m_a}{2 \pi f_a} \v{B}_0 \sin(m_a t) ,
\end{equation}
which in turn produces a magnetic field $\v{B}_a$ that satisfies $\v{\nabla} \times \v{B}_a = \v{j}_a$, and can be amplified with an LC circuit and then detected using a magnetometer \cite{Sikivie2014LC}. An analogous strategy has also been proposed for the detection of hidden photons \cite{Dobrich2014B}.

\subsubsection{Other proposals}
\label{Sec:other}
Ref.~\cite{Baker2012} proposes to modify existing microwave cavity searches for axions through the insertion of radio-frequency cavities into dipole magnets from particle accelerators, wiggler magnets developed for accelerator-based advanced light sources, and toroidal magnets similar to those used in particle physics detectors, while Ref.~\cite{Rybka2015} proposes to modify existing microwave cavity detectors using an open Fabry-P\'{e}rot resonator. Ref.~\cite{Ringwald2010} suggests to search for the electromagnetic radiation emitted by conducting surfaces when they are excited by ALPs (or hidden photons). Refs.~\cite{Beck2013,Beck2015} propose to search for a Shapiro step-like signal induced by axions in Josephson junctions. Ref.~\cite{Arvanitaki2014NMR} proposes to exploit nuclear magnetic resonance to search for axion-mediated forces, while Ref.~\cite{Sikivie2014Atomic} proposes to search for axion-induced atomic transitions in macroscopic samples. We also note that an external time-dependent magnetic field may enhance the local energy density stored in an ALP field by several orders of magnitude \cite{Arias2015enhance}, which may have application to laboratory axion searches.

\section{Variations of the fundamental constants of Nature}
\label{VFCs}
\subsection{Traditional observations and models}
The idea that the fundamental constants of Nature might vary with time can be traced as far back as the large numbers hypothesis of Dirac, who hypothesised that the gravitational constant $G$ might be proportional to the reciprocal of the age of the universe, $G \propto 1/t$ (which was later shown to be inconsistent with observations) \cite{Dirac1937,Dirac1938,Dirac1974}.
Since Dirac's initial hypothesis, a number of models, in which the fundamental constants vary with space and time, have been proposed and investigated, including Bekenstein models \cite{Bekenstein1982,Pospelov2002,SBM2002,Graham2013alpha}, string dilaton models \cite{Damour1994A,Damour1994B,Martins2015}, chameleon models \cite{Khoury2004}, and via quantum effects induced by cosmological renormalisation group flow \cite{Sola2005,Sola2009,Fritzsch2012,Markkanen2014} (see also the review~\cite{Uzan2011LRR} for an overview of other models). Anthropic arguments point out that `fine-tuning' of the fundamental constants is required for life to exist --- if the fundamental constants were even slightly different in our region of the Universe, then life could not have appeared. Variation of the fundamental constants of Nature across space provide a natural explanation to such `fine-tuning':~mankind simply appeared in a region of the Universe where the values of the fundamental constants are suitable for our existence.

Many laboratory, terrestrial and astrophysical searches for possible variations in the fundamental constants have been conducted. In any search for variations of the fundamental constants, the observable must be dimensionless (for otherwise, the observable would depend on the choice of units). Atomic transition frequencies are sensitive to variations in the electromagnetic fine-structure constant, $\alpha = e^2 / \hbar c$, and particle masses; for instance, the leading-order relativistic corrections to the hydrogenic Dirac energy levels scale as $\propto (Z\alpha)^2$. Atomic clock and spectroscopy measurements in the laboratory using a wide range of systems have provided some of the most precise limits on temporal drifts in $\alpha$ and particle masses to date \cite{Stein1974Cs-oscillator,Prestage1995H-alkalis,Dzuba1999A,Dzuba1999B,Sortais2001Rb-Cs,Bize2003Rb-Cs,Bize2003Hg+Cs,Fischer2004H-Cs,Peik2004Yb+-Cs,Tedesco2006,Budker2007Dy-Dy,Ye2008Sr-Cs,Wineland2008Al+Hg+,Leefer2013Dy-Dy,Peik2014Yb+Cs,Gill2014Yb+Yb+}. The most stringent laboratory limits on temporal variations of $\alpha$ come from Al$^+$/Hg$^+$ \cite{Wineland2008Al+Hg+} and Yb$^+$/Yb$^+$ \cite{Gill2014Yb+Yb+} clock comparison experiments:
\begin{equation}
\label{alpha_clock_limit}
|(d\alpha/dt)/\alpha| \lesssim 2 \times 10^{-17}~\textrm{year}^{-1} ,
\end{equation}
while the most stringent laboratory limits on temporal variations of the electron-to-proton mass ratio $m_e/m_p$ come from Yb$^+$/Cs clock comparison experiments \cite{Peik2014Yb+Cs}:
\begin{equation}
\label{me/mp_clock_limit}
|d(m_e/m_p)/(m_e/m_p)| \lesssim 2 \times 10^{-16}~\textrm{year}^{-1} .
\end{equation}
A variety of systems have been proposed to provide improved laboratory limits on temporal variations in the fundamental constants. 
Optical transitions in highly-charged ionic species that are near the crossing of different electronic configurations are very sensitive to variations in $\alpha$ \cite{Berengut2010HCI,Berengut2011HCI}.
Molecules, in which there is near cancellation between hyperfine and rotational intervals \cite{Flambaum2006molecules}, ground-state fine-structure and vibrational intervals \cite{Flambaum2007molecules}, and omega-type doubling and hyperfine intervals \cite{Flambaum2013molecules}, have enhanced sensitivity to $\alpha$, $m_e/m_p$ and the ratio of light quark masses to the QCD scale $m_q/\Lambda_{\textrm{QCD}}$. Molecular ions, in which transitions occur between nearly degenerate levels of different nature, have enhanced sensitivity to $\alpha$ and $m_e/m_p$ \cite{Pasteka2015mol-ion}.
Transitions in some nuclei, such as the ultraviolet transition between the ground and first excited states in the $^{229}$Th nucleus, have highly enhanced sensitivity to variations in $\alpha$ \cite{Flambaum2006Th}. Most recently, laser interferometers have been proposed to search for variations of $\alpha$ \cite{Stadnik2015laser}. The phase accumulated in an interferometer arm, $\Phi = \omega L / c$, changes if the fundamental constants change (the arm length $L = N a_B$, where $a_B$ is the Bohr radius, and atomic frequency $\omega$ both depend on the fundamental constants), according to $\delta \Phi \simeq \Phi ~ \delta \alpha / \alpha$, with a typical accumulated phase of $\Phi \sim 10^{11}$ in a single passage for an optical transition and $L = 4$ km. Multiple reflections enhance the accumulated phase and effects of variation of $\alpha$ by the factor $N_{\textrm{eff}} \sim 100$.

Stringent terrestrial limits on the temporal variation of $\alpha$ have been determined from the Oklo phenomenon. Roughly $1.8$ billion years ago, a uranium-rich natural nuclear reactor in Oklo went critical, consumed a portion of its fuel and then shut down several million years later. The isotopic ratio $^{149}$Sm/$^{147}$Sm (neither of which are fission products) in Oklo ores is much lower than in ordinary samarium, which can be interpretted as been due to the depletion of $^{149}$Sm by the capture of thermal neutrons by $^{149}$Sm: $n + ~^{149}\textrm{Sm} \to~^{150}\textrm{Sm} + \gamma$, which is dominated by the capture of a neutron with an energy of about 0.1 eV. This low-energy resonance arises due to the near cancellation between the electromagnetic and strong interactions, and the position of this resonance depends strongly on $\alpha$, from which the following limit on temporal variations of $\alpha$ is obtained \cite{Shlyakhter1976}:
\begin{equation}
\label{Oklo_limit}
|(d\alpha/dt)/\alpha| \lesssim 10^{-17}~\textrm{year}^{-1} .
\end{equation}

Shifts in quasar absorption spectral lines provide a powerful tool to search for spatial variations in $\alpha$ and $m_e/m_p$ \cite{Savedoff1956,Webb1999,Webb2001,Levshakov2003,Aracil2004,Flambaum2007NH3,Webb2011,Ubachs2015}. The most recent independent data samples from the VLT and Keck telescopes both indicate the presence of a spatial gradient in $\alpha$ \cite{Webb2011}:
\begin{equation}
\label{alpha_dipole}
\frac{\Delta \alpha}{\alpha} \simeq 10^{-16} ~\textrm{ly}^{-1} .
\end{equation}
A consequence of this astronomical result is that, since the solar system is moving along this spatial gradient, there should exist a corresponding temporal shift in $\alpha$ in Earth's frame of reference at the level $\delta \alpha / \alpha \sim 10^{-19}~\textrm{year}^{-1}$ \cite{Berengut2012test}. Finding this variation with laboratory experiments may independently corroborate the astronomical result. The dynamics of electron-proton recombination is governed by $\alpha$ and $m_e$. CMB measurements thus provide a means of probing possible variations of the fundamental constants, with sensitivities at the fractional level of $\sim 10^{-2} - 10^{-3}$ for variations of $\alpha$ and $m_e$ from the present-day values, respectively \cite{Hannestad1999CMB,Turner1999CMB,Martins2000CMB,Avelino2001CMB, Landau2001CMB,Weller2001CMB}. The primordial light elemental abundances, which are produced during BBN, are sensitive to changes in the fundamental constants, with sensitivities to temporal variations in the constants from the present-day values at the fractional level $\sim 10^{-2}$ \cite{Flambaum2002BBN,Cyburt2005BBN,Landau2006BBN,Wetterich2007BBN,Coc2007BBN,Berengut2010BBN,Bedaque2011BBN,Berengut2013BBN}.

The underlying cause of any possible variations in the fundamental constants is still an open question. Traditional dark energy-based models, which predict a cosmological evolution of the fundamental constants, invoke a (nearly) massless underlying field. DM-based models offer a more natural approach to producing variations of the fundamental constants of Nature, since the underlying DM fields do not necessarily need to be exceedingly light. In the remaining sections, we present an overview of the mechanisms through which slow `drifts', oscillating variations and transient-in-time variations of the fundamental constants may be induced by DM.

\subsection{Scalar interactions and constraints}
\label{Scalar+Astrophys}
Scalar fields may interact linearly with SM matter as follows
\begin{align}
\label{Scalar_couplings_lin}
\mathcal{L}_{\textrm{scalar}}^{\textrm{lin.}} = \frac{\phi}{\Lambda_\gamma} \frac{F_{\mu \nu} F^{\mu \nu}}{4} - \sum_{f} \frac{\phi}{\Lambda_f} m_f \bar{f} f + \sum_{V} \frac{\phi}{\Lambda_V} \frac{M_V^2}{2} V_\nu V^\nu ,
\end{align}
where the first term represents the coupling of the scalar field to the electromagnetic field tensor $F$, the second term represents the coupling of the scalar field to the fermion bilinears $\bar{f} f$, while the third term represents the coupling of the scalar field to the massive vector boson wavefunctions. 
The various $\Lambda_X$ that appear in Eq.~(\ref{Scalar_couplings_lin}) are the respective new physics energy scales (analogous to the axion decay constant $f_a$ in Eq.~(\ref{Axion_couplings})), which are independent of one another and are known to be very large energy scales from fifth-force searches (the exchange of an ultralight scalar particle between two SM fermions produces an attractive Yukawa potential, $V(r) = - m_f^2 / 4 \pi \Lambda_f^2 \cdot e^{-m_\phi r} / r$), which include lunar laser ranging \cite{Turyshev2004,Turyshev2012} and the E\"{o}tWash experiment \cite{Adelberger2008,Adelberger2012}.
The most stringent bounds for the scalar masses $m_\phi \lesssim 10^{-14}$ eV are from the E\"{o}tWash experiment \cite{Adelberger2008,Adelberger2012}:
\begin{align}
\label{scalar_linear_bounds_Eot}
\Lambda_\gamma \gtrsim 3 \times 10^{22}~\textrm{GeV},~ \Lambda_e \gtrsim 2 \times 10^{21}~\textrm{GeV}, \notag \\
\hat{\Lambda}_q = (m_d+m_u) \Lambda_d \Lambda_u / (m_d \Lambda_u + m_u \Lambda_d) \gtrsim 5 \times 10^{23}~\textrm{GeV} .
\end{align}
Stellar energy-loss arguments applied to the bremmstrahlung processes $e^- + ~^4\textrm{He} \to e^- + ~^4\textrm{He} + \phi$ give the following limit on the linear coupling of a scalar field to the electron \cite{Raffelt1999}:
\begin{equation}
\label{scalar_e_limit}
\Lambda_e \gtrsim 4 \times 10^{10}~\textrm{GeV}
\end{equation}
while similar arguments applied to the Compton-like process $\gamma + ~^4\textrm{He} \to \phi + ~^4\textrm{He}$ give the following limit on the linear coupling of a scalar field to the proton \cite{Raffelt1999}:
\begin{equation}
\label{scalar_p_limit}
\Lambda_p \gtrsim 2 \times 10^{10}~\textrm{GeV} .
\end{equation}

Likewise, (pseudo)scalar DM may also interact quadratically with SM matter as follows
\begin{align}
\label{Scalar_couplings_quad}
\mathcal{L}_{\textrm{scalar}}^{\textrm{quad.}} = \frac{\phi^2}{(\Lambda'_\gamma)^2} \frac{F_{\mu \nu} F^{\mu \nu}}{4} - \sum_{f} \frac{\phi^2}{(\Lambda'_f)^2} m_f \bar{f} f + \sum_{V} \frac{\phi^2}{(\Lambda'_V)^2} \frac{M_V^2}{2} V_\nu V^\nu ,
\end{align}
where the $\Lambda'_X$ are much less severely constrained than the corresponding $\Lambda_X$, from astrophysical observations and fifth-force searches. Stellar energy-loss arguments applied to the photon pair annihilation process $\gamma + \gamma \to \phi + \phi$ constrain the quadratic coupling of a scalar field to the photon \cite{Olive2008}:
\begin{equation}
\label{scalar_gamma_limit_QUAD}
\Lambda'_\gamma \gtrsim 3 \times 10^{3}~\textrm{GeV} ,
\end{equation}
while similar arguments applied to the nucleon bremmstrahlung process $N + N \to N + N + \phi +\phi$ give the following limit on the quadratic coupling of a scalar field to the proton \cite{Olive2008}:
\begin{equation}
\label{scalar_proton_limit_QUAD}
\Lambda'_p \gtrsim 15 \times 10^{3}~\textrm{GeV} ,
\end{equation}
and similarly for the bremmstrahlung process $e^- + (A,Z) \to e^{-} + (A,Z) + \phi + \phi$, which constraints the quadratic coupling of a scalar field to the electron as follows \cite{Olive2008}:
\begin{equation}
\label{scalar_proton_limit_QUAD}
\Lambda'_e \gtrsim 3 \times 10^{3}~\textrm{GeV} .
\end{equation}
For the quadratic couplings of Eq.~(\ref{Scalar_couplings_quad}), a fifth-force is produced in the leading order by the exchange of a pair of $\phi$-quanta, which generates a less efficient $V(r) \simeq - m_f^2/ 64 \pi^3 (\Lambda'_f)^4 \cdot 1 / r^3$ attractive potential, instead of the usual Yukawa potential in the case of linear couplings. The resulting constraints from fifth-force experiments are hence weakened significantly \cite{Olive2008}:
\begin{equation}
\label{scalar_proton_limit_QUAD_ff}
\Lambda'_p \gtrsim 2 \times 10^{3}~\textrm{GeV} ,
\end{equation}
which hold for $m_\phi \lesssim 10^{-4}$ eV \cite{Adelberger2007+}.

\subsection{Oscillating variations of the fundamental constants}
\label{Osc_VFCs}

\begin{figure}[h!]
\begin{center}
\includegraphics[width=10cm]{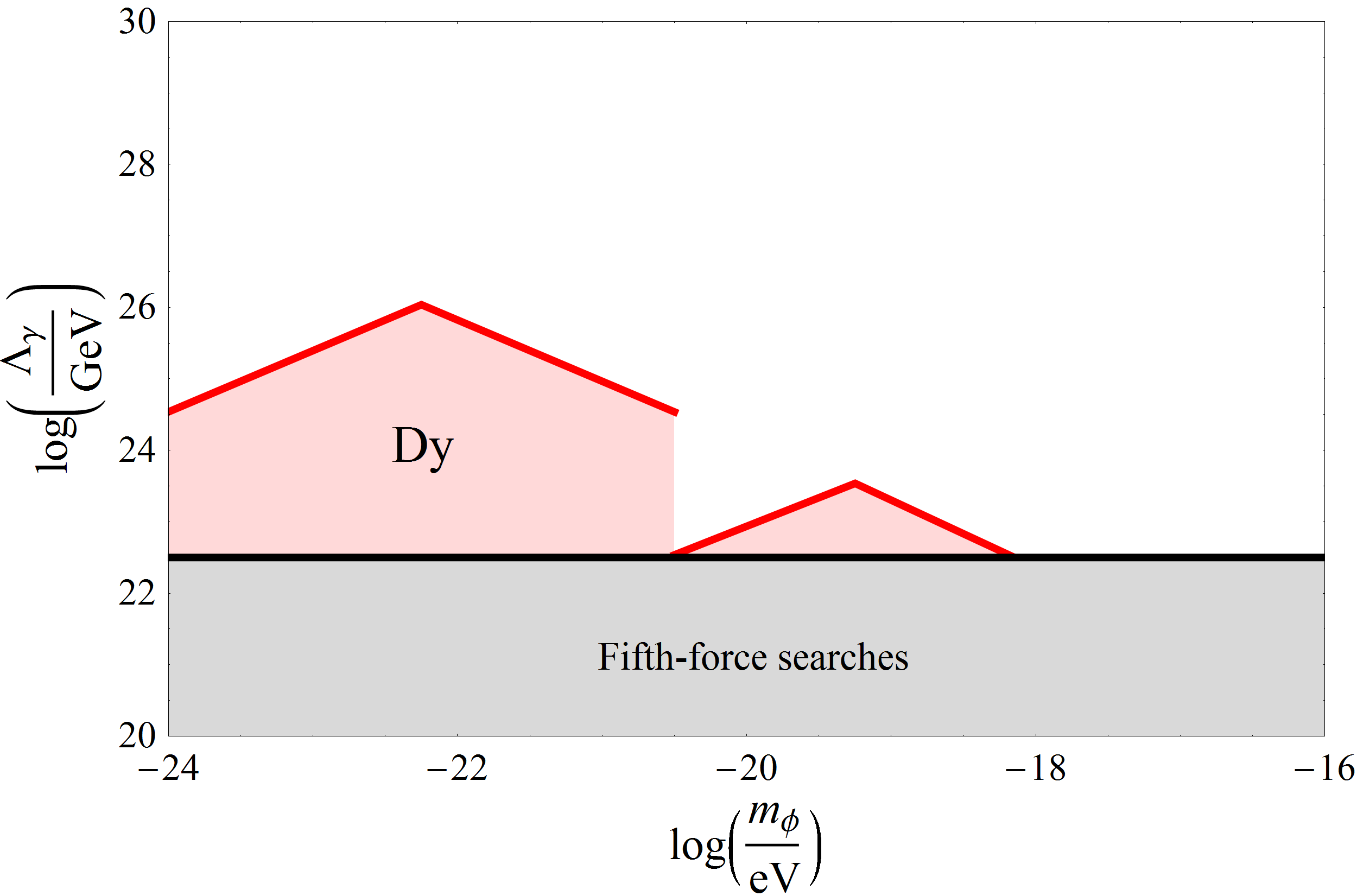}
\caption{Region of scalar dark matter parameter space ruled out for the linear interaction of $\phi$ with the photon. 
Region below red line corresponds to constraints from atomic dysprosium spectroscopy measurements \cite{Budker2015scalar}. 
Region below black line corresponds to constraints from fifth-force experimental searches \cite{Adelberger2008,Adelberger2012}.} 
\label{fig:Lambda_gamma_linear_space}
\end{center}
\end{figure}

The couplings in Eq.~(\ref{Scalar_couplings_lin}) alter the electromagnetic fine-structure constant $\alpha$ and particle masses as follows
\begin{align}
\label{Lin_constants}
\alpha \to \frac{\alpha}{1-\phi/\Lambda_\gamma} \simeq \alpha \left[1 +\frac{\phi}{\Lambda_\gamma} \right] , \frac{\delta m_f}{m_f} = \frac{\phi}{\Lambda_f} , \frac{\delta M_V}{M_V} = \frac{\phi}{\Lambda_V} ,
\end{align}
and, likewise, the couplings in Eq.~(\ref{Scalar_couplings_quad}) alter $\alpha$ and the particle masses as follows
\begin{align}
\label{Quad_constants}
\alpha \to \frac{\alpha}{1-\phi^2/(\Lambda'_\gamma)^2} \simeq \alpha \left[1 +\frac{\phi^2}{(\Lambda'_\gamma)^2} \right] , \frac{\delta m_f}{m_f} = \frac{\phi^2}{(\Lambda'_f)^2} , \frac{\delta M_V}{M_V} = \frac{\phi^2}{(\Lambda'_V)^2} .
\end{align}
The couplings of an oscillating scalar field to the SM fields via the linear interactions in Eq.~(\ref{Scalar_couplings_lin}) \cite{Arvanitaki2015scalar,Stadnik2015laser,Stadnik2015BBN} and via the quadratic interactions in Eq.~(\ref{Scalar_couplings_quad}) \cite{Stadnik2015laser,Stadnik2015BBN,Stadnik2015DM-VFCs} produce oscillating variations of the fundamental constants, which can be sought for with high-precision terrestrial experiments involving atomic clocks \cite{Arvanitaki2015scalar,Stadnik2015BBN,Stadnik2015DM-VFCs} and laser interferometers \cite{Stadnik2015laser}. A multitude of atomic, highly-charged ionic, molecular and nuclear systems can be used in clock-based searches, see the reviews \cite{Flambaum2008clock_review,Flambaum2014HCI_review} for summaries of the possible systems. The first laboratory clock-based search for oscillating variations of $\alpha$ has very recently been completed \cite{Budker2015scalar}, and the results have been used to place stringent constraints on the photon interaction parameters $\Lambda_\gamma$ \cite{Budker2015scalar} (Fig.~\ref{fig:Lambda_gamma_linear_space}) and $\Lambda'_\gamma$ \cite{Stadnik2015BBN,Stadnik2015DM-VFCs} (Fig.~\ref{fig:Lambda_gamma_quadratic_space_+CMB}) for the range of scalar DM masses $10^{-24}~\textrm{eV} \lesssim m_\phi \lesssim 10^{-16}~\textrm{eV}$.

\subsection{`Slow' drifts of the fundamental constants}
\label{Slow_VFCs}
\begin{figure}[h!]
\begin{center}
\includegraphics[width=10cm]{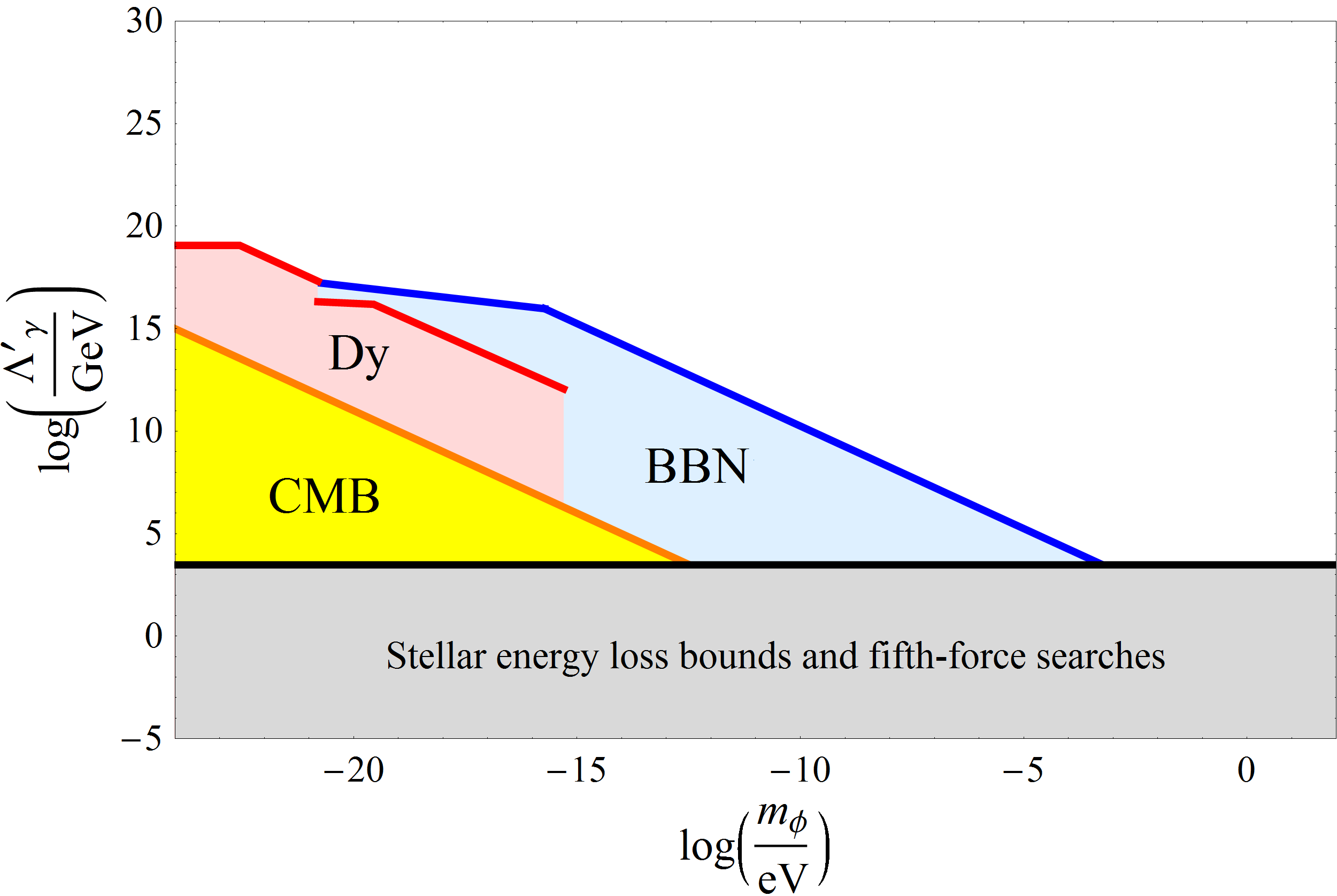}
\caption{Region of scalar dark matter parameter space ruled out for the quadratic interaction of $\phi$ with the photon. 
Region below red line corresponds to constraints from atomic dysprosium spectroscopy measurements \cite{Stadnik2015BBN,Stadnik2015DM-VFCs}. 
Region below yellow line corresponds to constraints from CMB measurements \cite{Stadnik2015DM-VFCs}. 
Region below blue line corresponds to constraints from comparison of measurements and SM calculations of the ratio $Q_{np}/T_F$ \cite{Stadnik2015BBN,Stadnik2015DM-VFCs}.
Region below black line corresponds to constraints from stellar energy loss bounds and fifth-force experimental searches \cite{Olive2008}.} 
\label{fig:Lambda_gamma_quadratic_space_+CMB}
\end{center}
\end{figure}

The coupling of an oscillating scalar field to the SM fields via the quadratic couplings in Eq.~(\ref{Scalar_couplings_quad}) produces not only oscillating variations of the fundamental constants, but also `slow' drifts of the fundamental constants through the $\left< \phi^2 \right> = \phi_0^2 / 2$ terms in Eq.~(\ref{Quad_constants}) \cite{Stadnik2015BBN,Stadnik2015DM-VFCs}, which is related to the local ambient density of non-relativistic scalar DM via the relation $\rho_{\textrm{scalar}} \simeq m_\phi^2 \phi_0^2 / 2$. 
The dynamics of electron-proton recombination is governed by $\alpha$ and $m_e$. CMB measurements constrain the quadratic interactions of $\phi$ with the photon and electron as follows \cite{Stadnik2015DM-VFCs}: 
\begin{align}
\label{CMB_quad_constraints}
\Lambda'_{\gamma,e} \gtrsim \frac{1~\textrm{eV}^2}{m_\phi} .
\end{align}
Changes in the fundamental constants during and prior to BBN alter the primordial abundances of the light elements. The observed and calculated (within the SM) ratio of the neutron-proton mass difference to freeze-out temperature at the time of weak interaction freeze-out ($t_{\textrm{F}} \approx 1.1$ s), which determines the abundance of neutrons available for BBN (with the vast majority of these neutrons ultimately being locked up in $^{4}$He), constrain the quadratic interactions of $\phi$ with the photon, light quarks and massive vector bosons as follows \cite{Stadnik2015BBN,Stadnik2015DM-VFCs}:
\begin{align}
\label{constraints_WIFO_3a}
\frac{1}{m_\phi^2}   \left\{ \frac{0.08 }{(\Lambda'_\gamma)^2} + \frac{1.59}{m_d-m_u} \left[\frac{ m_d}{(\Lambda'_d)^2} - \frac{ m_u}{(\Lambda'_u)^2} \right] + \frac{3.32 }{(\Lambda'_W)^2}  - \frac{4.65 }{(\Lambda'_Z)^2} \right\} \notag \\
 \simeq  (1.0 \pm 2.5) \times 10^{-20} ~\textrm{eV}^{-4} ,
\end{align}
when the scalar field is oscillating at $t_{\textrm{F}} \approx 1.1$ s ($m_\phi \gg 10^{-16}$ eV). 
When the scalar field had not yet begun to oscillate at $t_{\textrm{F}} \approx 1.1$ s ($m_\phi \ll 10^{-16}$ eV), the corresponding constraints on the quadratic interactions of $\phi$ are \cite{Stadnik2015BBN,Stadnik2015DM-VFCs}:
\begin{align}
\label{constraints_WIFO_3a++}
\frac{1}{m_\phi^2} \left(\frac{m_\phi}{3 \times 10^{-16} ~\textrm{eV}}\right)^{3/2}  &\left[ \frac{0.08 }{(\Lambda'_\gamma)^2} + \frac{1.59}{m_d-m_u} \left(\frac{ m_d}{(\Lambda'_d)^2} - \frac{ m_u}{(\Lambda'_u)^2} \right)    + \frac{3.32 }{(\Lambda'_W)^2}  - \frac{4.65 }{(\Lambda'_Z)^2} \right] \notag \\
&\simeq  (0.5 \pm 1.3) \times 10^{-20} ~\textrm{eV}^{-4} ,
\end{align}
while the constraints on the linear interactions of $\phi$ are \cite{Stadnik2015BBN}:
\begin{align}
\label{constraints_WIFO_3a++L}
\frac{1}{m_\phi} \left(\frac{m_\phi}{3 \times 10^{-16} ~\textrm{eV}}\right)^{3/4}  &\left[ \frac{0.08 }{\Lambda_\gamma} + \frac{1.59}{m_d-m_u} \left(\frac{ m_d}{\Lambda_d} - \frac{ m_u}{\Lambda_u} \right)     + \frac{3.32 }{\Lambda_W}  - \frac{4.65 }{\Lambda_Z} \right] \notag \\
&\simeq  (0.4 \pm 1.0) \times 10^{-11} ~\textrm{eV}^{-2} .
\end{align}
The constraints on the photon interaction parameter $\Lambda'_\gamma$ from CMB and BBN measurements are shown in Fig.~\ref{fig:Lambda_gamma_quadratic_space_+CMB}.

\subsection{Transient-in-time variations of the fundamental constants}
\label{Transient_VFCs}
The coupling of a scalar DM field that comprises a topological defect with SM fields via either of the couplings in Eqs.~(\ref{Scalar_couplings_lin}) or (\ref{Scalar_couplings_quad}) alters the fundamental constants and particle masses inside the defect, giving rise to local transient-in-time variations as a defect temporarily passes through this region \cite{Derevianko2014,Stadnik2014defects}. These transient-in-time variations can be sought for using a global network of detectors, including atomic clocks \cite{Derevianko2014} and laser interferometers \cite{Stadnik2015laser}, as well as a network of pulsars \cite{Stadnik2014defects}, such as the international pulsar timing array \cite{PTA2010}, or binary pulsar systems \cite{Stadnik2015defects_reply}. For sufficiently non-adiabatic passage of a defect (a relatively thin and/or rapidly travelling defect) through a pulsar, a topological defect may trigger a pulsar glitch \cite{Stadnik2014defects,Stadnik2015defects_reply}. Numerous pulsar glitches have already been observed (see e.g.~Ref.~\cite{Lyne2006_pulsar_book} for an overview), but their underlying cause is still debated (see e.g.~Ref.~\cite{Haskell2015_glitch_review} for a review).

\section{Outlook}
\label{Outlook}
Exciting developments in dark matter searches are expected over the next few years. 
Effects that are linear in the interaction constant between dark matter and ordinary matter provide strong motivation for a new generation of searches for ultralight axion and scalar dark matter. 
The first such laboratory search for ultralight scalar dark matter by means of atomic spectroscopy measurements in dysprosium has already been completed, placing new constraints on the linear and quadratic interactions of ultralight scalar dark matter with the photon that surpass previous constraints by many orders of magnitude.
A number of other laboratory searches for ultralight dark matter using atomic and solid-state magnetometry, atomic clocks, interferometry, torsion pendula and ultracold neutrons are either already in progress or planned to commence in the near future.
These experiments are expected to yield limits on the interaction parameters of ultralight dark matter with ordinary matter that are many orders of magnitude better than existing limits and, more importantly, offer reinvigorated hope for the unambiguous direct detection of dark matter.

\section*{Acknowledgements}
We would like to thank Maxim Yu.~Khlopov for the invitation to write this book review.
This work was supported by the Australian Research Council. 
V.~V.~F.~is grateful to the Mainz Institute for Theoretical Physics (MITP) for its hospitality and support.

\bibliographystyle{ws-rv-van}
\bibliography{ws-rv-sample}
\printindex                         
\end{document}